\documentclass{article}
\usepackage{amsmath}
\pdfoutput=1
\usepackage{graphicx}
\usepackage{natbib}

\begin{document}

\begin{center}
{\LARGE
Protein Folding: A Perspective From Statistical Physics}

\vspace{0.5cm}

Jinzhi Lei$^{1}$, and Kerson Huang$^{2,3}$

$^{1}$ Zhou Pei-Yuan Center for Applied Mathematics, Tsinghua University, Beijing, 100084, China

$^{2}$  Physics Department, Massachusetts Institute of Technology, Cambridge,  MA02139, USA

$^{3}$  Institute of Advanced Studies, Nanyang Technological University, Singapore, 639673

\end{center}

\vspace{1.0cm}

\textbf{ABSTRACT} In this paper, we introduce an approach to the protein folding problem from the point of view of statistical physics. Protein folding is a stochastic process by which a polypeptide folds into its characteristic and functional 3D structure from random coil. The process involves an intricate interplay between global geometry and local structure, and each protein seems to present special problems. The first part of this chapter contains a concise discussion on kinetics versus thermodynamics in protein folding, and introduce the statistical physics basis of protein folding. In the second part, we introduce CSAW (conditioned self-avoiding walk), a model of protein folding that combines the features of self-avoiding walk (SAW) and the Monte Carlo method. In this model, the unfolded protein chain is treated as a random coil described by SAW. Folding is induced by hydrophobic forces and other interactions, such as hydrogen bonding, which can be taken into account by imposing conditions on SAW. Conceptually, the mathematical basis is a generalized Langevin equation. Despite the simplicity, the model provides clues to study the universal aspects while we overlook details and concentrate only on a few general properties. To illustrate the flexibility and capabilities of the model, we consider several examples, including helix formation, elastic properties, and the transition in the folding of myoglobin. From the CSAW simulation and physical arguments, we find a universal elastic energy for proteins, which depends only on the radius of gyration $R_{g}$ and the residue number $N$. The elastic energy gives rise to scaling laws $R_{g}\sim N^{\nu }$ in different regions with exponents $\nu =3/5,3/7,2/5$, consistent with the observed unfolded stage, pre-globule, and molten globule, respectively. These results indicate that CSAW can serve as a theoretical laboratory to study universal principles in protein folding.

\newpage

\section{Introduction}

\index{protein folding} In spite of considerable efforts in the past
decades, the protein folding problem is still unsolved and remains one of
the most basic intellectual challenges in molecular biology %
\citep{BrandenTooze, Shakhnovich:2006p1580, Zhang:2008p6799}. The principle
through which the amino acid sequence determines the native structure, as
well as the dynamics of the process, remain open questions.

In the past 36 years, thinking in the protein folding field has being strongly influenced by 
\index{thermodynamic hypothesis} Christian Anfinsen's \textquotedblleft thermodynamic hypothesis\textquotedblright. This hypothesis states that the native
state of a protein is the one with the lowest Gibbs free energy, and is determined by the totality of interatomic interactions of the amino-acid 
sequence, in a given environment \citep{Anfinsen:1973}. Predicting the 3D structure of a protein from its amino-acid sequence is one of the most important goals  in the protein folding problem \citep{MoPeJuFi95, ZhangSkolnick, Zhang:2008p6799}.

Despite this widely established thermodynamic hypothesis 
\citep{Dill90, KimBaldwin90}, recent experiments have suggested that there may be exceptions, especially for larger and complex proteins \citep{BakerAgard:1994, Baker, LazaridisKarplus, Shakhnovich:2006p1580}. For these proteins, the native conformations correspond to the kinetically most accessible state (kinetic control), instead of the most stable one (thermodynamic control) \citep{LazaridisKarplus}. This would be the case when the barriers between conformations are too high to be ergodic kinetically, and the protein ends up at a state of local minimum energy that is accessible in available time.

\index{molecular dynamics} Molecular dynamics (MD) is an invaluable tool
with which to study protein folding dynamics \textit{in vitro} %
\citep{DayDaggett, ScheragaKhaliliLiwo}. MD solves the Newtonian equations
of the motion of all atoms in a protein on a computer, using appropriate
interatomic potentials. To describe the solvent, one includes thousands of
water molecules explicitly, treating all atoms in the water on the same
footing as those on the protein chain. For the integration algorithm to be
stable, the time step $\Delta t$ must be an order of magnitude smaller than
the fastest motions of the system, typically the vibration of a
covalent bond, whose period is the order of 10 fs. Thus, the time step is of
the order of 1 fs \citep{McCGelKar, pearlman95}. Not surprisingly, such an extravagant use of computing power is so inefficient that one can follow the folding process only to about a microsecond  \citep{DuanKollman, JangKimShinPak}, whereas the folding of a real protein takes from seconds to minutes. The use of reduced models (meso-scopic, coarse-grained, implicit-solvent) with physically based potentials is a reasonable trade-off between computational cost and accuracy 
\citep{Levitt, KazLiwSch02, KamLiwSch03, Nielsen04, Huang:2007, Sun07}. 

Proteins with different amino acid sequences can invoke quite different
folding mechanisms \citep{DaggettFersht}, while proteins with high sequence
similarity can end up with very different folds \citep{Alexander05, He05}.
Nevertheless, universal aspects do emerge, if one overlooks details and
concentrates only on a few general properties. These will be the main subject in this chapter.

We are concerned here with a perspective of protein folding from the
point of view of statistical physics \citep{Huang:2005}. After all, the
protein is a chain molecule immersed in water, and, like all physical
systems, will tend towards thermodynamic equilibrium with the environment.
The process is stochastic, involving an intricate interplay between global
geometry and local structure. Our goal is to design a model amenable to computer simulation in a reasonable time, and to investigate the physical principles of protein folding, in particular the relative importance of various interactions.

We treat the protein as a molecular chain performing Brownian motion in
water, regarded as a medium exerting random forces on the chain, with the
concomitant energy dissipation. In addition, we include regular (non-random)
interactions within the chain, as well as between the chain and the medium.
Protein folding is a stochastic process of conformation changes, and must
be analyzed in terms of a statistical ensemble, but not of a single pathway. In this perspective, we study protein folding by defining a  transition probabilities between the conformations, which are
described through a Monte Carlo type model.

The unfolded chain is assumed to be a random coil described by SAW
(self-avoiding walk), as suggested by Flory some time ago \citep{Flory}.
That is, links in the chain correspond to successive random steps, in
which the chain is prohibited from revisiting an occupied position.

We model the protein chain in 3D space, keeping only degrees of freedom
relevant to folding, which we take to be the torsional angles between
successive links. In a computer simulation, we first generate an ensemble
of SAW's, and then choose a sub-ensemble through a Monte Carlo method, which
generates a canonical ensemble with respect to a Hamiltonian that specifies
the interactions. We call the model 
\index{CSAW} CSAW (conditioned self-avoiding walk) \citep{Huang:2007}.
Mathematically speaking, it is based on a Langevin equation describing the
Brownian motion of a chain with interactions. There seems little doubt that
such an equation does describe a protein molecule in water, for it is just
Newton's equation with the environment treated as a stochastic medium. 

Two types of interactions are included in our initial formulation:
\begin{itemize}
\item the hydrophobic interaction with the medium, which causes the chain to
fold;
\item hydrogen bonding within the chain, which leads to secondary structures.
\end{itemize}
The model can be implemented efficiently, and is
flexible enough to be used as a theoretical laboratory. In our initial
studies, we keep only the two interactions listed above, in order to describe the folding dynamics qualitatively. Other
interactions, such as electrostatic and van der Waals interactions, can be
add as refinements.

Both CSAW and MD are based on Newtonian mechanics, and differ only in the
idealization of the system. In CSAW we replace the thousands of water
molecules used in MD by a stochastic medium---the heat reservoir of
statistical mechanics. With this simplification, we consider the main
interactions---hydrophobic interaction and heat dissipation, and ignore
inessential degrees of freedom, such as small vibration in the length and
angles of the chemical bonds in the protein chain. The advantages of this
idealization are that we
\begin{itemize}
\item avoid squandering computer power on irrelevant calculations;
\item gain a better physical understanding of the folding process.
\end{itemize}

After a brief review of the basics of protein folding and stochastic
processes, we shall describe the model in more detail, and illustrate its
use through examples involving realistic protein fragments. We will
demonstrate helix formation, elastic properties, and folding stages.

Our results indicate that the CSAW model can describe qualitative features
in the folding of simple proteins, and provide physical insight on the
mechanisms of protein folding.

\section{Protein basics}

\subsection{The protein chain}

The 
\index{protein} protein is a polypeptide chain consists of a sequence of
units or 
\index{residue}\textquotedblleft residues\textquotedblright, which are amino acids chosen
from a pool of 20. This sequence is called the 
\index{protein structure!primary structure} \textit{primary structure}.
Along the protein chain, the amino acids are connected by 
\index{peptide bond}{peptide bonds} between the C atom of one residue and
the nitrogen atom of the next. The center of each amino acid is a carbon
called $\mathrm{C}_{\alpha }$. The amino acids differ from each other only
in the side chains connected to the $\mathrm{C}_{\alpha }$'s. The side chains
can be grouped to different classes by their hydrophobicity, charge, and
polarity \citep{BrandenTooze}.

\begin{figure}[tbph]
\centering
\includegraphics[width=10cm]{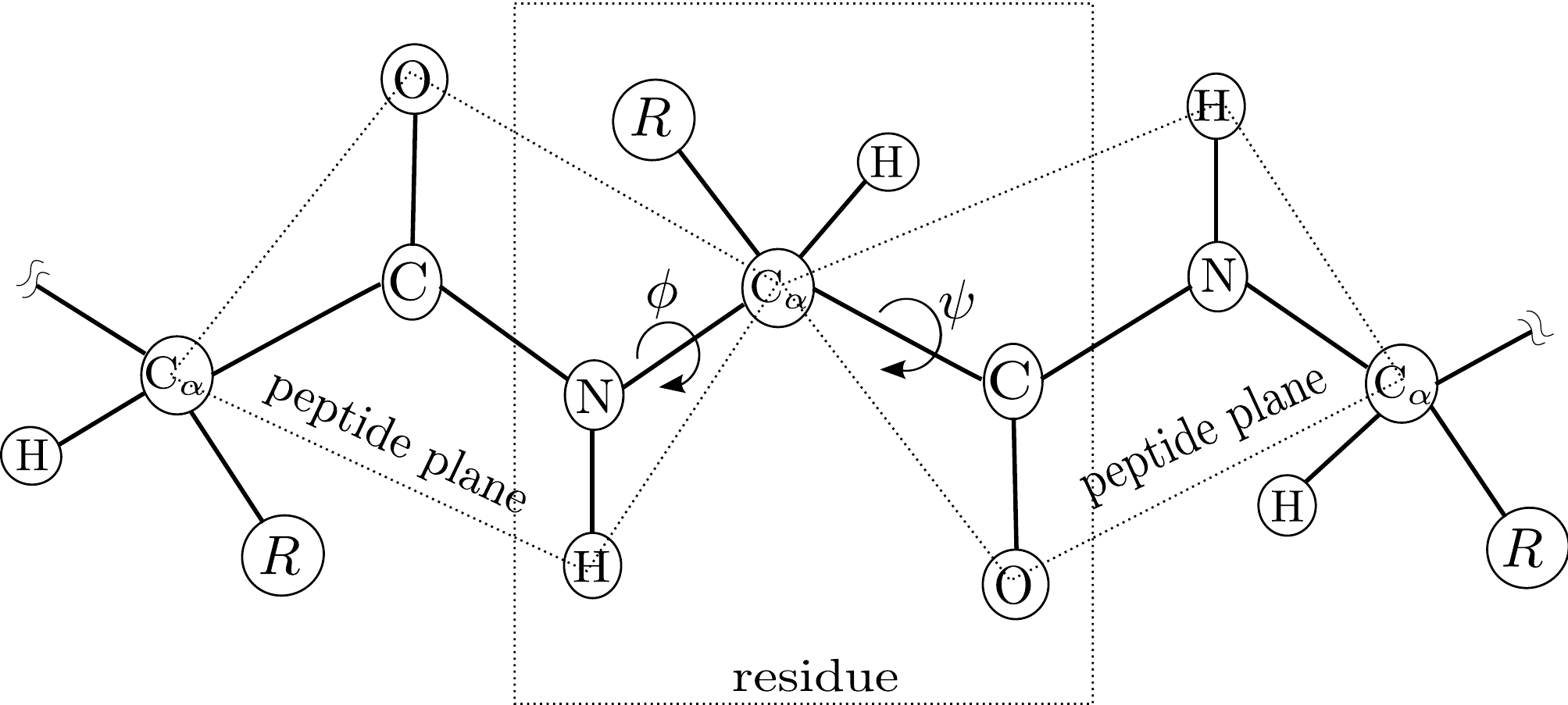}
\caption{Schematic representation of the protein chain. The center of a residue
is a carbon atom labeled $\protect\alpha $. The residues are connected by
the peptide bonds between the C atom of one residue and the nitrogen atom of
the next. All atoms connected to the peptide bond lie on a plane, the
peptide plane. The only degrees of freedom we consider are the torsion
angles $\phi, \psi $ that specify the relative orientation
of successive peptide planes. Residues can differ only in the side chains
labeled $R$, chosen from a pool of twenty.}
\label{fig:residue}
\end{figure}

A polypeptide chain can also be viewed as repeating peptide units that
connect one $\mathrm{C}_{\alpha }$ atom to the next one along the backbone
(Figure \ref{fig:residue}). All atoms in such a unit are fixed in a plane,
the 
\index{peptide plane} peptide plane, with the bond lengths and bond angles
very nearly the same in all units in all proteins (Table \ref{tab:const}).

The relative orientation of successive units is determined by the two 
\index{torsion angle} torsion angles (aka dihedral angle) $\phi $ and $\psi $%
, as schematically illustrated in Figure \ref{fig:residue}. The torsion
angles $\phi $ and $\psi $ are rotation angles about the $\mathrm{N}$-$\mathrm{C}
_{\alpha }$ and $\mathrm{C}_{\alpha }$-$\mathrm{C}$ bonds, respectively, the
positive sense being defined according to the right-hand rule. By
definition, the flat chain, with all backbone atoms lying in a plane,
corresponds to $\phi =\psi =180^{\circ }$ for all residues.

When small vibrations in bond lengths and bond angles are neglected, the
torsion angles are the only degrees of freedom. For our
purpose, therefore, a protein of $N$ residues has $2(N-1)$ degrees of
freedom, i.e., the angle pairs $(\phi _{n},\psi _{n})_{n=1}^{N}$, where $n$
labels the residue\footnote{The angles $\phi _{1}$ and $\psi _{N}$ are not relevant}. The angle pair $(\phi _{n},\psi _{n})$ can take values only from sterically allowed regions in the 
\index{Ramachandran plot} \textquotedblleft Ramachandran plot\textquotedblright.

\subsection{Secondary structures}

The natural proteins have evolved through natural selection to perform
specific biological functions, which depend on their 3D structures or 
\index{conformation} conformations. In general, the 3D structure in the
native state 
\index{protein structure!native structure} contains one or more domains,
each of which is made up of ordered 
\index{protein structure!secondary structure} secondary structures, such as 
\index{secondary structure ! $\alpha $ helix} alpha ($\alpha $) helices or 
\index{secondary structure ! $\beta $ sheet} beta ($\beta $) sheets, and
disordered 
\index{secondary structure ! loop} loops. When the secondary structures are
blurred over, one sees a global shape called the 
\index{protein structure!tertiary structure} tertiary structure. 

The alpha helix is the most common form of secondary structure in proteins.
It consists of a stretch of consecutive residues all having torsion angle pairs $(\phi ,\psi
)\approx (-60^{\circ },-50^{\circ })$ \citep{Pauling51}. The helix has 3.6
residues per turn. A hydrogen bond connects the (CO) in residue $n$ to the
(NH) in residue $n+4$. There are variations in which residue $n$ is bonded
to $n+3$ or $n+5$, instead of $n+4$.

The beta sheet is another form of secondary structure, in which different \index{secondary structure ! $\beta $ strand} 
beta strands are connected laterally by five or more hydrogen bonds, forming
a twisted pleated sheet. Here, a beta strand refers to a stretch of amino
acids whose peptide backbones are almost fully extended, with $(\phi ,\psi
)\approx (-135^{\circ },135^{\circ })$. In the Ramachandran plot this
corresponds to a broad region in the upper left quadrant.

\subsection{Hydrogen bonds}

A 
\index{hydrogen bond} hydrogen bond (H-bond) refers to the sharing of the H
atom between the groups (N-H) and (O=C). It is formed when (a) the distance
between H and O fall within certain limits, and (b) the chemical bonds in
the two groups are antiparallel, within a tolerance. The bond is partly of
electrostatic nature, and partly covalent. It is weaker than the usual
covalent or ionic bond, but stronger than the van der Waals interaction.

The number of hydrogen bonds in a typical protein is very large, and they
are important for the mechanical stability of a conformation. The (NH) and
(CO) groups on the backbone tend to form hydrogen bonds with each another.
Hydrophilic side chains can also form hydrogen bonds. They are usually located on
the protein surface, and bond with the water solvent.

\subsection{Hydrophobic effect}

In the water molecule $\mathrm{H}_{2}\mathrm{O}$, all the atoms can form hydrogen bonds with
another water molecule. Thus, water in bulk consists of a network of
hydrogen bonds. In ice, they form a 3D lattice in which each $\mathrm{H}_{2}\mathrm{O}$ has 4 nearest neighbors.

In the medium of protein folding, liquid water, there is a fluctuating
network of H-bonds with an average life time of $9.5\mathrm{ps}$.hydrophilic That is,
the bonding partners change on a time scale of $10\mathrm{ps}=10^{-11}%
\mathrm{sec}$ \citep{Garrett}. A foreign molecule introduced into water
would disrupt the network, unless it can participate in hydrogen bonding. If
it can form H-bond with water, it is said to be \textquotedblleft soluble\textquotedblright, or \textquotedblleft 
hydrophilic\textquotedblright, and will be received by water molecules as
one of their kind. Otherwise it is unwelcome, and said to be
\textquotedblleft insoluble\textquotedblright, or \textquotedblleft hydrophobic\textquotedblright. Protein side chains can be hydrophobic or
hydrophilic\footnote{The hydrophobic side chains include: Ala (A), Val (V), Leu (L),
Ile (I), Phe (F), Pro (P), Met (M), Trp (W). The hydrophilic side chains include: Asp (D), Glu (E), Lys (K), Arg (R), Ser (S), Thr (T), Cys
(C), Asn (N), Gln (Q), His (H), Tyr (Y), Gly (G) \citep{Garrett}.}; but the (CO) and (NH) groups on the backbone are hydrophilic.

When immerse in water, the protein chain rearranges its conformation in
order to shield the hydrophobic residues from water. In effect, the water
networks squeezes the protein into shape. This is called the 
\index{hydrophobic effect} \textquotedblleft hydrophobic
effect\textquotedblright. A \textquotedblleft frustration\textquotedblright arises in the rearrangement process, when
burying a hydrophobic side chain drags part of the backbone into the interior
of the protein. Since the backbone is always hydrophilic, this robs it of the
chance to bond with water. The frustration is resolved by the formation of
secondary structures, which \textquotedblleft use up\textquotedblright the hydrogen bonds
internally \citep{BrandenTooze}. The folded chain reverts to a random coil
when the temperature becomes too high, or when the pH of the solution
becomes acidic.

\subsection{Statistical nature of the folding process}

We have to distinguish between protein assembly inside a living cell (%
\textit{in vivo}) and folding in a test tube (\textit{in vitro}). The former
process takes place within factory molecules called ribosomes, and generally
needs the assistance of \textquotedblleft chaperon\textquotedblright\
molecules to prevent premature folding. In the latter, the molecules freely
fold or unfold, reversibly, depending on the pH and the temperature.

We deal only with folding \textit{in vitro}, in which many 
protein molecules undergo the folding process independently, and they do not
fold in unison. We are thus dealing with an ensemble of protein molecules,
in which definite fractions exist in various stages of folding at any given
time. The Langevin equation naturally describes the time evolution of such
an ensemble. Behavior of individual molecules fluctuate from the average,
even after the ensemble has reached equilibrium. In a macroscopic body
containing the order of $10^{23}$ atoms, such fluctuations are unobservable
small. For a protein with no more than a few thousand atoms, however, these
fluctuations are pronounced.

To understand the physical basis of protein folding, we
should analyze the time evolution of the ensemble from the perspective of
statistical physics, instead of the evolution of a single pathway.

\subsection{Folding stages}

A typical folding process consists of a fast initial collapse into an
intermediate state call 
\index{protein folding!molten globule}the \textquotedblleft molten
globule\textquotedblright . The latter takes a relatively long time to
undergo fine adjustments to reach the native state. The collapse time is
generally less than 300$\mu $s \citep{Akiya02, Uzawa04, Kimura05}, while 
the molten globule can last as long as 10 minutes. In some proteins
there is evidence for a 
\index{protein folding!pre-globule} pre-globule stage %
\citep{Uversky96, Uversky02}.

In the 
\index{protein folding!denature state} 
\index{protein folding!unfolded state}unfolded state, a protein chain is
extended and flexible, whereas it is globular and compact in the 
\index{protein folding!folded state} folded, 
\index{protein folding!native state} native state. The molten globule is as
compact as the folded state, and possesses most of the secondary structures.
The pre-globule is unstructured and less compact.

An overall characteristic of the protein structure is the 
\index{radius of gyration} radius of gyration $R_{g}$, the root-mean-square
separation between residues, which can be measured by small angle X-ray
scattering \citep{Chu74, Berne76, Doi86}. Stages in the folding process can
be characterized by 
\index{scaling law} scaling relations of the form $R_{g}\sim N^{\nu }$. The
unfolded protein chain, which is akin to a homopolymer, has index $\nu =3/5$
according to Flory's model \citep{Flory}. The observed index of pre-globule is $\nu =0.411\pm 0.016\approx 3/7$ \citep{Uversky02}. The native
state proteins has a smaller index $\nu =2/5$ 
\citep{Arteca94,
Arteca95, Arteca96,Hong2009}.

To anticipate results described in more detail later, we propose the 
\index{Protein folding!folding stages}folding stages depicted in Figure \ref%
{fig:stages}, with the indicated time scales and the scaling relations $%
R_{g}\sim N^{\nu }.$%
\begin{figure}[tbph]
\centering
\includegraphics[width=10cm]{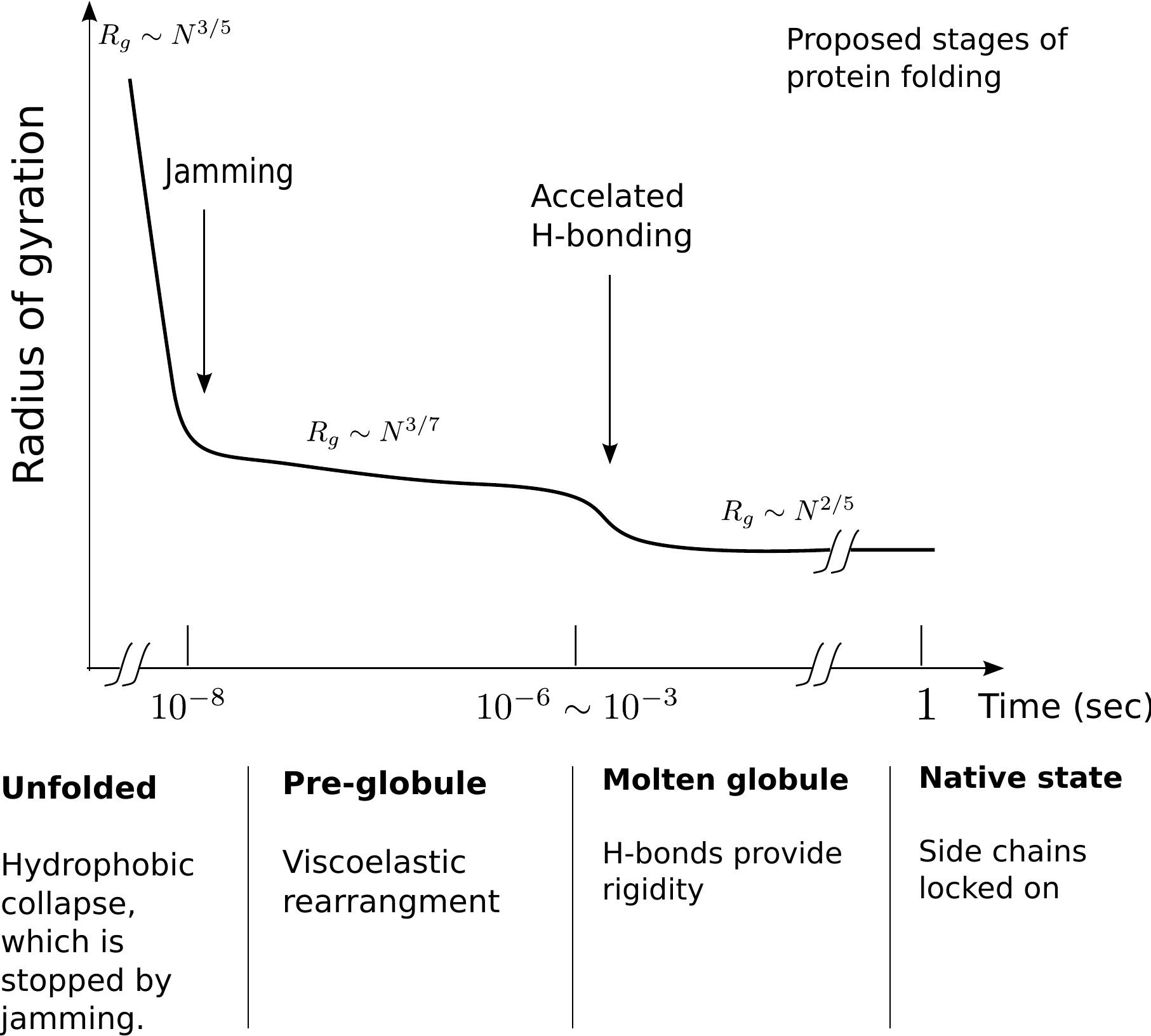}
\caption{Suggested stages in protein folding based on physical arguments and
computer simulations to be discussed later. Each stages is characterized by a scaling law of the
form $R_{g}\sim N^{\protect\nu }$, where $R_{g}$ is the radius of gyration
and $N$ is the number of residues. While the existence of these stages
should be a universal property of proteins, the visibility of the pre-globule
stage is not universal, because its lifetime depends on details of
short-ranged interactions.}
\label{fig:stages}
\end{figure}

\section{Physical basics}

\subsection{Stochastic process}

Protein folding is a 
\index{stochastic process} stochastic process \citep{Mich2006, Dill2008},
involving random forces ever present in the formation and destruction of
hydrogen bonds, thermal fluctuations in covalent bonds and bond angles, etc.
An individual protein molecule does not have a fixed conformation,
even in the equilibrium state; it is characterized by a probability
distribution about average conformations.

The folding process is described by the time evolution of a
probability distribution $P(X,t)$, where $X$ denotes a conformation (or
configuration). We can model it through the master equation \citep{vanKampen}
\begin{equation}
\dfrac{\partial P(X,t)}{\partial t}=\int \{W(X|X^{\prime },t)P(X^{\prime
},t)-W(X^{\prime }|X,t)P(X,t)\}dX^{\prime }  \label{eq:master}
\end{equation}%
Here $W(X|X^{\prime },t)$ is the 
\index{transition probability} transition probability, at time $t$, per unit
time, from the conformation $X^{\prime }$ to $X$.

For simplicity, we assume that protein folding is a 
\index{Markov process} Markov process, which means the transition
probability depends only on its present state, and not on any past state.
That is, there is no memory of the previous history. The transition
probability therefore is independent of the time $t$, and rewritten $%
W(X|X^{\prime })$. This transition probability is all we need to describe
the folding process. The equilibrium state is characterized
by the probability distribution $P(X)$ satisfying 
\begin{equation}
\int \{W(X|X^{\prime })P(X^{\prime })-W(X^{\prime }|X)P(X)\}dX^{\prime }=0
\label{eq:equi}
\end{equation}%
for all conformations $X$.

The Markov assumption is the simplest one can make, and it leads to the
well-established Gaussian distribution, the \textquotedblleft bell curve"
applicable in many statistical problems. In the macroscopic world, there is
of course a persistence of memory. The only question is how important it is.
The study of non-Markovian processes leads to the field of \textquotedblleft
complexity\textquotedblright, upon which we will not tread.

\subsection{Brownian motion}

In the 
\index{Brownian motion}Brownian motion of a single particle suspended in a
medium, one can physically verify that the Markov hypothesis is justified.
The transition probability can be derived analytically as follows.

Consider 1D Brownian motion in a potential well $U(x)$. The position $x(t)$
is a stochastic variable described by the Langevin equation %
\citep{Huang2001} 
\begin{equation}
m%
\ddot{x}=F(t)-\gamma \,\dot{x}-U^{\prime }(x),  \label{eq:bm}
\end{equation}%
where $U^{\prime }(x)=dU(x)/dx$. The force on the particle by the medium is
split into two parts: a random fluctuation $F(t)$ and a damping force $%
-\gamma \,\dot{x}$. The random force is a member of a statistical ensemble
with the properties 
\begin{equation}
\begin{array}{c}
\langle F(t)\rangle =0 \\ 
\langle F(t)F(t^{\prime })\rangle =c_{0}\delta (t-t^{\prime })%
\end{array}%
\end{equation}%
where the brackets $\langle \ \rangle $ denote ensemble average. The two
forces are not independent, but related through the 
\index{fluctuation-dissipation theorem} fluctuation-dissipation theorem: 
\begin{equation}
\dfrac{c_{0}}{2\gamma }=k_{B}T
\end{equation}%
where $k_{B}=1.38\times 10^{-23}J/K$ is 
\index{Boltzmann's constant} Boltzmann's constant, and $T$ is the absolute
temperature.

In an overdamped medium, we can omit the particle acceleration, and rewrite %
\eqref{eq:bm} as a stochastic differential equation 
\begin{equation}
\gamma \,%
\dot{x}=F(t)-U^{\prime }(x).
\end{equation}%
In a coarse time scale, the transition $\Delta x=x^{\prime }-x$ in $\Delta t$%
, starting from $x$, satisfies 
\begin{equation}
\dfrac{\langle \Delta x\rangle _{x}}{\Delta t}=-\gamma ^{-1}\,U^{\prime
}(x),\quad \dfrac{\langle (\Delta x)^{2}\rangle _{x}}{\Delta t}=2D+(\gamma
^{-1}U^{\prime}(x))^2\Delta t,  \label{eq:bmd}
\end{equation}%
where $D=k_{B}T/\gamma $ is the diffusion constant by 
\index{Einstein's relation} Einstein's relation.

Let $W(\Delta x,\Delta t;x)$ to be the probability of a transition from $x$
to $x^{\prime }=x+\Delta x$ in $\Delta t$, starting from $x$. From 
\eqref{eq:bmd} we can deduce  
\begin{equation}
W(\Delta x,\Delta t;x)=%
\dfrac{1}{\sqrt{4\pi \,D\,\Delta t}}\exp \left[ -\dfrac{\left( \Delta
x+\gamma ^{-1}\,U^{\prime }(x)\right) ^{2}}{4D\,\Delta t}\right] .
\label{eq:tpbm0}
\end{equation}%
This Gaussian distribution verifies the central limit theorem, according to
which the probability of the sum of a large number of any independent random
variables is Gaussian \citep{Huang2001}.

From \eqref{eq:tpbm0}, the transition probability is obtained by setting $%
\Delta t=1$: 
\begin{equation}
W(x^{\prime }|x)=\dfrac{1}{\sqrt{4\pi D}}\exp \left[ -\dfrac{\left( \Delta
x+\gamma ^{-1}\,U^{\prime }(x)\right) ^{2}}{4D}\right] .  \label{eq:tpbm}
\end{equation}%
This enables us to write down the
equation for the probability distribution $P(x,t)$. When the potential is
absent, the equation can be solved exactly, and also simulated by 
\index{random walk} random walk. Both methods lead to diffusion, in which
the position has a Gaussian distribution with variance $%
\sqrt{2Dt}$ \citep{vanKampen}.

When there is an external force, we may not be able to solve the equation
analytically, but we can still simulate it on a computer by 
\index{conditioned random walk} conditioned random walk as discussed below.

\subsection{Conditioned random walk}

In the conditioned random walk, we first generate a trial step by random
walk, but accept it with a \ certain probability. To obtain the probability
of a transition $\Delta x$ over a time interval $\Delta t$, we integrate $%
W(\Delta x,\Delta t;x)$ over $\Delta t,$ and get 
\begin{equation}
W(\Delta x;x)=\int_{0}^{\infty }W(\Delta x,\Delta t;x)d\Delta t=c\exp \left[
-%
\dfrac{\Delta U(x)}{2D\gamma }\left( 1+\dfrac{\Delta U(x)}{|\Delta U(x)|}%
\right) \right] ,  \label{eq:proM}
\end{equation}%
Here $\Delta U(x)\approx U^{\prime }(x)\,\Delta x$, and $c=|1/(\gamma
^{-1}\,U^{\prime }(x))|=|\Delta t/\langle \Delta x\rangle _{x}|$. The
exponential in \eqref{eq:proM} gives following Metropolis criteria:
\begin{itemize}
\item if $\Delta U\leq 0$, accept it;
\item if $\Delta U>0$, accept it with probability $\exp (-\Delta U/k_{B}T)$.
\end{itemize}
The last condition allows for acceptance even if the energy
increases, and this simulates thermal fluctuations.

For simplicity, we have illustrated the method in the overdamped
approximation; but the results hold in general.

We can also integrate the equation \eqref{eq:bm} directly as a stochastic
differential equations, as an alternative to conditioned random walk. The
equivalence of these two methods is illustrated by an example in the
appendix of \cite{LeiHuang:2008}.

\section{CSAW model}

\subsection{Model description}

In protein folding, we are dealing with the Brownian motion of a
non-overlapping chain with interactions. To follow the time development, we begin with an unfolded chain modeled by self-avoiding walk
(SAW), and take interactions into account through conditions imposed on
updates. The resulting model is called CSAW (conditioned self-avoid walk).

We can generate a SAW representing the unfolded protein chain by the 
\index{pivot algorithm} pivot algorithm, as follows \citep{Li1995, Kennedy02}%
. Choose an initial self-avoiding walk in 3D continuous space, and hold one
end of the chain fixed.
\begin{enumerate}
\item Choose an arbitrary point on the chain as pivot.
\item Rotate the end portion of the chain rigidly about the pivot (by
changing the torsional angles at the pivot point).
\item If this does not result in any overlap, accept the conformation;
otherwise repeat the procedure.
\end{enumerate}

By the method, we can generate a uniform ergodic ensemble of SAW's, which
simulates a Langevin equation of the form 
\begin{equation}
m_k 
\ddot{\mathbf{x}}_k = \mathbf{F}_k(t) - \gamma_k\, \dot{\mathbf{x}}_k + 
\mathbf{V}_k(\mathbf{x}),\quad (k = 1,\cdots,N)
\end{equation}
where the subscripts $k$ labels the residues along the chain. The terms $%
\mathbf{V}_k$ denote the regular (non-random) forces that maintain the rigid
bonds between successive residues, the bond angles, and that prohibit the
residues from overlapping one another.

We now add other regular forces $\mathbf{G}_{k}=-\nabla _{\mathbf{x}_{k}}E(%
\mathbf{x})$, where $E$ is the potential to be detail latter. Analogy to the
above discussion, it gives an acceptance probability
\begin{equation}
W(\Delta \mathbf{x}; \mathbf{x}) = \left\{\begin{array}{ll}
1, &\mathrm{if}\ \Delta E \leq 0\\
\exp (-\Delta E/2k_{B}T)& \mathrm{if}\ \Delta E > 0
\end{array}\right.
\end{equation}
of a random walk from $\mathbf{x}$ to $\mathbf{x}' = \mathbf{x} + \Delta \mathbf{x}$. Treating this
force via the Metropolis method results in CSAW, which simulates a
generalized Langevin equation indicated in the following:%
\begin{equation}
m_{k}\ddot{\mathbf{x}}_{k}\ =\underset{\text{Treat via SAW}}{\mathbf{F}%
_{k}(t)-\gamma _{k}\,\dot{\mathbf{x}}_{k}+\mathrm{V}_{k}(\mathbf{x})}\quad -%
\underset{\text{Treat via Metropolis}}{\nabla _{\mathbf{x}_{k}}E(\mathbf{x})}
\end{equation}
Now we shall specify the potential explicitly.

\subsection{Implementation of CSAW}

To reiterate, the system under consideration is a sequences of residues $(\mathrm{NH}$-$\mathrm{C}_{\alpha }R\mathrm{H}$-$\mathrm{CO})$ connected by
peptides, as shown in Figure \ref{fig:residue}. The residues can differ from
one another only through the side chains $R$ attached to the $\mathrm{C}%
_{\alpha }$ atom, and there are 20 of them to choose from. There are O and H
atoms attached to each residue. The atoms between two successive $\mathrm{C}%
_{\alpha }$ atoms are fixed in a plane, with fixed bond lengths and bond
angles as given in Table. \ref{tab:const}. The degrees of freedom of the
system are the pairs of torsional angles $(\phi _{n},\psi _{n})$
specifying the relative orientation of two successive peptide planes.

\begin{table}[hbtp]
\caption{Bond lengths and bond angles used in the CSAW model.}
\label{tab:const}
\begin{center}
\begin{tabular}{|l|l||l|l|}
\hline
Bond & Value ($A$) & Angle & Value ($^\circ$)$^{(c)}$ \\ \hline
$\mathrm{C}_\alpha$-$\mathrm{N}$ & $1.46$ & $\mathrm{N}$-$\mathrm{C}_\alpha$-$\mathrm{C}$ & $111.0$ \\ 
$\mathrm{C}_\alpha$-$\mathrm{C}$ & $1.51$ & $R$-$\mathrm{C}_\alpha$-$\mathrm{C}$ & 
$101.1$ \\ 
$\mathrm{C}_\alpha$-$R$ & $1.53^{(a)}$ & $R$-$\mathrm{C}_\alpha$-$\mathrm{N}$ & $%
109.6$ \\ 
$\mathrm{C}_\alpha$-$\mathrm{H}$ & $1.00$ & $\mathrm{H}$-$\mathrm{C}_\alpha$-$\mathrm{C}$ & $101.1$ \\ 
$\mathrm{C}$-$\mathrm{N}$ & $1.33$ & $\mathrm{C}_\alpha$-$\mathrm{C}$-$\mathrm{N}$
& $114.0$ \\ 
$\mathrm{N}$-$\mathrm{H}$ & $1.00$ & $\mathrm{C}$-$\mathrm{N}$-$\mathrm{C}_\alpha$
& $123.0$ \\ 
$\mathrm{C}$-$\mathrm{O}$ & $1.24$ & $\mathrm{H}$-$\mathrm{N}$-$\mathrm{C}$ & $%
123.0$ \\ 
$\mathrm{N}$-$\mathrm{CH}_2$ & $1.48^{(b)}$ & $\mathrm{H}$-$\mathrm{N}$-$\mathrm{C}%
_\alpha$ & $114.0$ \\ 
&  & $\mathrm{O}$-$\mathrm{C}$-$\mathrm{N}$ & $125.0$ \\ 
&  & $\mathrm{O}$-$\mathrm{C}$-$\mathrm{C}_\alpha$ & $121.0$ \\ \hline
\end{tabular}%
\par
\begin{minipage}{8cm}
$^{(a)}$ Replaced by $\mathrm{C}_\alpha$-$\mathrm{H}$ for glycine.\\
$^{(b)}$ The bond length $\mathrm{N}$-$\mathrm{CH}_2$ at proline.\\
$^{(c)}$ Same values for glycine and proline.
\end{minipage}
\end{center}
\end{table}

For simplicity we treat all atoms, as well as the side chains, as hard
spheres. More realistic representation can be implement if desired. The
hard sphere sizes refer to van der Waals radii (Table \ref{tab:rwd}). Two
atoms (not neighbors along the chain) are \textquotedblleft in contact" if
their centers are separated by a distance less than the summation of their
van der Waals radii. They are regarded as overlapping if the distance is
less than a factor $\sigma (<1)$ \footnote{%
We use $\sigma =0.6$ in our studies.} times the contact distance. We have
to treat the residues glycine and proline as special cases, since the
side chain of glycine contains only one H atom, while the $\mathrm{N}$-$%
\mathrm{H}$ group in proline is replaced by $\mathrm{N}$-$\mathrm{CH}_{2}$
(Table \ref{tab:const}).

Hydrophobic interaction and hydrogen bonding are the dominant interactions
governing the protein folding and maintaining the stability of the folded
state. They are introduced in the energy function $E$ in the form 
\begin{equation}
E=-g_{1}K_{1}-g_{2}K_{2}  \label{eq:energy}
\end{equation}%
where $K_{1}$ is the total contact number of all hydrophobic residues, and $%
K_{2}$ the total number of internal hydrogen bonds. The parameters $%
g_{i}(i=1,2)$ are constants, regarded in this model as universal for all
proteins.

\begin{figure}[htbp]
\centering
\begin{minipage}[b]{5cm}
\begin{center}
\includegraphics[width=4.0cm]{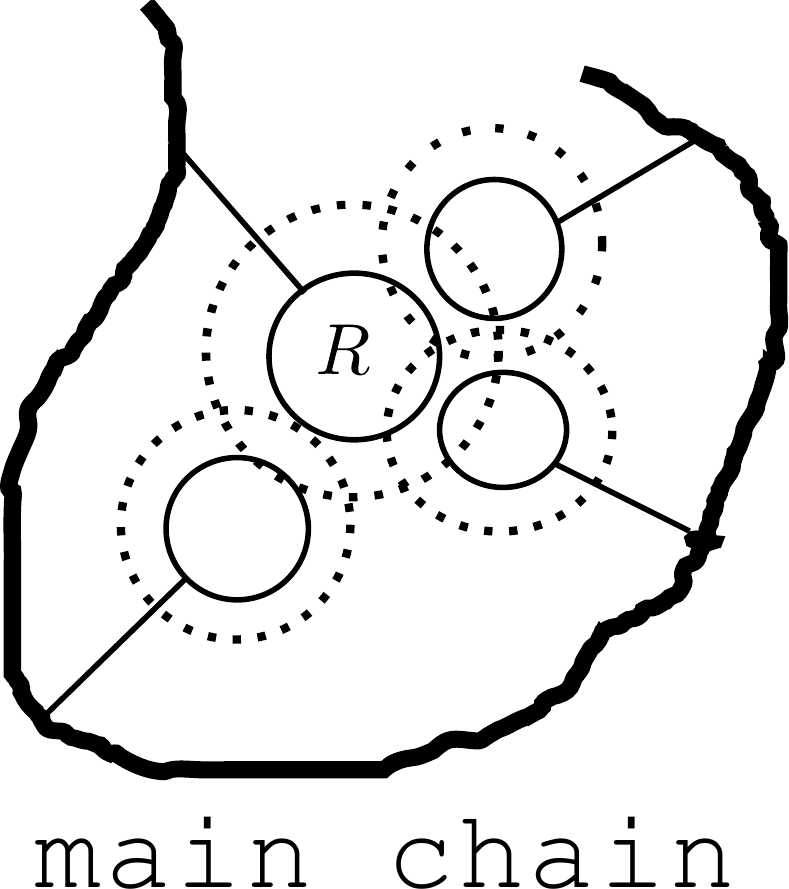}

\vspace{0.2cm}
(a)
\end{center}
\end{minipage}\hspace{1cm}%
\begin{minipage}[b]{5cm}
\begin{center}
\includegraphics[width=4.5cm]{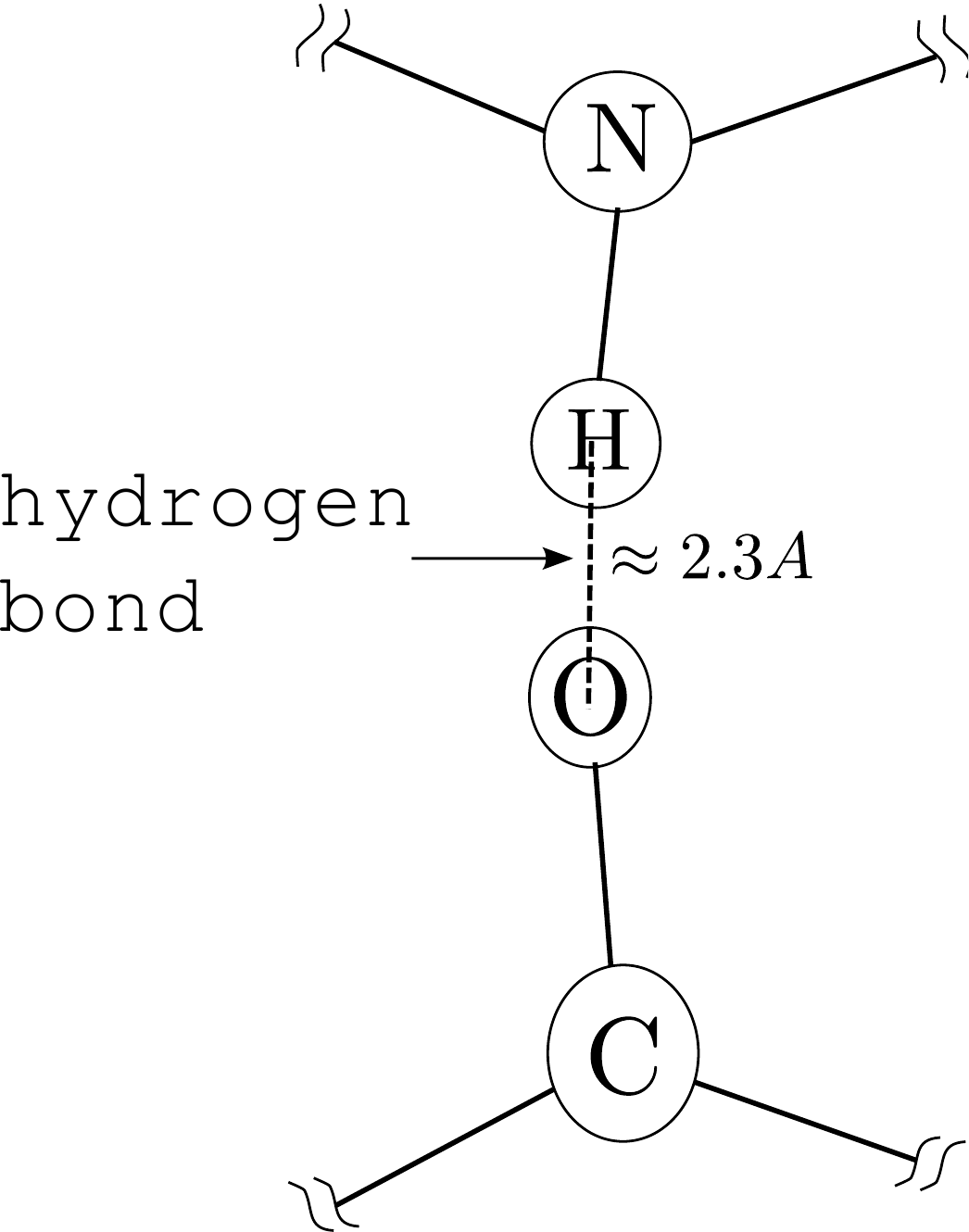}

\vspace{0.2cm}
(b)
\end{center}
\end{minipage}
\caption{(a) The hydrophobic residue illustrated here has three contact
neighbors. The permanent neighbors along the chain are not counted. Solid
and dotted lines correspond to hard sphere and van der Waals radii,
respectively. (b) Hydrogen bonding occurs between O and H on the main chain,
from different residues.}
\label{fig:contact}
\end{figure}

The first term in $E$ expresses the hydrophobic effect, which is
proportional to the number of water molecules surrounding the hydrophobic
side chains, i.e., 
\[
\mathrm{(Maximum\ No.)}-\mathrm{(Total\ hydrophobic\ contact\ No.)}. 
\]%
The maximum number contributes a constant to the total energy and can be
omitted. The contact number of a residue is the number of atoms touching its
side chain (not counting the two permanent nearest neighbors along the
chain). This is illustrated in Figure \ref{fig:contact}(a).

\begin{table}[hbtp]
\caption{van der Waals radius (A) of the atoms}
\label{tab:rwd}
\begin{center}
\begin{tabular}{ccccc}
\hline\hline
N & C & R & H & O \\ \hline
$1.55$ & $1.70$ & $1.95^{(a)}$ & $1.20$ & $1.50$ \\ \hline\hline
\end{tabular}%
\par
\begin{minipage}{8cm}
$^{(a)}$ Refer the value for $\mathrm{CH}_3$, same for $\mathrm{CH}_2$ in proline.
\end{minipage}
\end{center}
\end{table}

The contact number measures how well a hydrophobic residue is being shielded
from the medium. When two hydrophobic residues are in contact, the total
contact number increase by $2$, and this induces an effective attraction
between hydrophobic residues, as in the simple HP model \citep{Lau90}.

The second term in $E$ describes internal hydrogen bonding. As illustrated
in Figure \ref{fig:contact}(b), an internal hydrogen bond is deemed to have
formed between O and H from different residues when
\begin{enumerate}
\item the distance between O and H is 2.3A, within given tolerance; and
\item The bonds C=O and N-H are antiparallel, within given tolerance.
\end{enumerate}

Only the combinations $g_{1}/2k_{B}T$ and $g_{2}/2k_{B}T$ appear in the
Metropolis method. They are treated as adjustable parameters. To study
the folding process at different temperatures, we define a program
temperature $T^{\ast }=k_{B}T/g_{2}$, and use $T^{\ast }$ and $g_{1}^{\ast
}=g_{1}/g_{2}$ as independent parameters.  

Note that $E$ only includes the potential energy. We leave out the kinetic
energy because it contributes only a constant factor to the conformational probability of the ensemble, provided the masses involved are constants
independent of momenta.

Because we are using the torsion angles as generalized coordinates, the
canonically conjugate momenta are such that the effective masses depend on
the coordinates, and thus not constants. However, independent studies have
shown that the mass variations are generally less than 1\%, and thus may be
ignored \citep{Leong10}.

\section{CSAW results}

We summarize results from previous runs of CSAW. 

\subsection{Test runs}

During the testing stage of CSAW, we run a minimal program as described in 
\cite{Huang:2007}. For a chain of 30 residues, the main findings are the
following:
\begin{itemize}
\item Under the hydrophobic forces alone, without hydrogen bonding, the
chain folds into a reproducible shape. This shows that the hydrophobic
effect along can produce tertiary structure. There is no secondary structure
in this case. The chain rapidly collapses to the final structure without
passing through an intermediate state.
\item When there is no hydrophobic force and the interaction consists purely
of hydrogen bonding, the chain rapidly folds into one long alpha helix. When
both hydrophobic force and hydrogen bonding are taken into account,
secondary structure emerges.
\item The folding process exhibits two-stage behavior, with a fast collapse
followed by slow \textquotedblleft annealing\textquotedblright , in
qualitative agreement with experiments.
\end{itemize}

\subsection{Energy landscape}

As a more realistic example, we fold Chignolin, a synthetic peptide of 10
residues whose native state is a $\beta $-hairpin \citep{Huang:2007}. From an
ensemble of 100 folding paths, starting with the random coil, we construct a
view of  the energy landscape, and exhibit the \textquotedblleft folding funnel\textquotedblright. We can also
examine individual paths, which are quite different from one another. A
common characteristic is that the path gets trapped in some pocket of states
for a relatively long time, and suddenly break out, only to be trapped again---a behavior reminiscent of the Levy flight. 

\subsection{Dynamics of helix formation}

To determine the CSAW parameters, we consider 
\index{polyalanine}Polyalanine ($\mathrm{Ala}_{20}$), a protein fragment of
20 identical amino acids alanine, which is hydrophobic. This protein has
been studied by many authors, both \textit{in vitro} and \textit{in silico},
and the native state is known to be a single alpha helix 
\citep{Liu:1998, Smith2001,
Shen:2005}. In the CSAW model, the helix emerges from an initial
random coil \citep{LeiHuang:2008}. We run multiple trial simulations with
different parameters $g_{1}^*$ and $T^*$ in order to find the optimal
values, with the results
\begin{equation}
g_1^*=0.05,\quad T^* = 0.2.
\end{equation}%
According to the philosophy of the model, this should be universal for all
proteins. But, since only the hydrophobic interaction and hydrogen bonding are
considered, we expect that there will be variations from protein to protein. 

The study reveals some dynamical aspects of helix formation. Starting from a
random coil, helical structure is nucleated at two locations along the
protein chain. The nuclei grow and eventually merge into the single helix
that is the native state. We obtain rate constants by fitting simulation
data to exponential functions. Using experimental information, we calibrate
one Metropolis MC step in CSAW as%
\begin{equation}
1\text{ MC step }\approx 10^{-12}\text{ s}.
\end{equation}

\section{Elastic energy of proteins}

\subsection{Flory analysis}

Stages in protein folding can be characterized by scaling relations of the
form%
\begin{equation}
R_{g}\sim N^{\nu }
\end{equation}
between the radius of gyration $R_{g}$ and the residue number $N$. This
exponent should be a universal feature independent of specific protein
sequence. A combination of theoretical modeling and experimental data
suggests that there are three stages with $\nu =3/5,3/7,2/5$, which may be
identified as the unfolded, pre-globule, and molten globule stage,
respectively \citep{Arteca94,
Uversky02, Hong2009}. We shall look into the existence and universality of
these stages via CSAW simulations. But first, a theoretical orientation.

The unfolded protein chain is akin to a homopolymer, for which the $3/5$
scaling law is known for long time from 
\index{Flory's theory} Flory's SAW (self-avoiding walk) model \citep{Flory}.
To derive the exponent, Flory models the free energy by%
\begin{equation}
F_{%
\mathrm{Flory}}=\frac{aR_{g}^{2}}{N}+\frac{bN^{2}}{R_{g}^{D}},
\end{equation}%
where $a$ and $b$ are temperature-dependent coefficients, and $D$ is the
spatial dimension. The first term is the free energy of stretching, in the
form of Hooke's law. It is assumed that $R_{g}^{2}$ scales like $N$, because
that is the behavior in random walk. The second term arises from the
excluded-volume effect, and is proportional to $N$ times the density.
Minimizing the energy with respect to $R_{g}$ leads to $\nu =3/(D+2)$, which
give $\nu =3/5$ for $D=3$.

The exponent 3/5 coincides with the Kolmogorov exponent in the energy
spectrum of turbulence. This is no accident, for turbulence consists of a
tangle of vortex lines, which can be modeled by SAW \citep{Huang:2005}.   

Hong and Lei \citep{Hong2009} generalize Flory's stretching free energy by
replacing $R_{g}^{2}/N$ with $R_{g}^{2}/N^{(2/\alpha )-1}:$%
\begin{equation}
F_{\mathrm{Hong-Lei}}=\frac{aR_{g}^{2}}{N^{\left( 2/\alpha \right) -1}}+\frac{%
bN^{2}}{R_{g}^{D}},
\end{equation}%
where $\alpha $ is the fractal dimension of the conformation. Minimizing this free energy leads to $\nu =(\alpha
+2)/[\alpha (D+2)]$, which for $D=3$ reduces to%
\begin{equation}
\nu =\frac{\alpha +2}{5\alpha }.
\end{equation}

Flory's $3/5$ law is recovered by setting the fractal dimension $\alpha =1$.

For proteins in the native state, the fractal dimension is $\alpha =2,$ as
can be deduced from analysis of data in the 
\index{PDB} PBD (Protein Data Bank). This leads to $\nu =2/5$. At the
free-energy minimum, the native-state free energy obeys Hooke's law, and
scales as%
\begin{equation}
F_{0}\sim R_{g}^{2}\sim N^{4/5}.  \label{4/5}
\end{equation}

These exponents will guide our analysis of data from CSAW simulations of
protein folding.

\subsection{Elastic energy}

What we can directly calculate via CSAW is not the free energy but an
 \textquotedblleft elastic energy\textquotedblright. 

We use CSAW to generate an ensemble of folding paths for a number of
proteins, and examine the average potential energy as a function of $R_{g}$ %
\citep{LeiHuang:2009}. We analyze the simulation data in terms of an elastic
potential energy, guided by the physical picture that the folding dynamics
is dominated by the twin actions of the hydrophobic force and hydrogen
bonding.

Five proteins are chosen for simulation, with residue numbers ranging from $%
N=20$ to $330$ (Table \ref{Table:proteins}). For each protein, the simulation generates an ensemble of $%
1.2\times 10^{4}$ conformations.

\begin{table}[h] 
\centering%
\caption{Proteins simulated. $N$ = number of residues, $h$ = fraction of hydrophobic residues.}
\begin{tabular}{|l|l|l|l|l|}
\hline
Protein name & ID & $N$ & $h$ & Structure \\ 
\hline\hline
Polyalanine & ala20 & $20$ & $1.000$ & 1 alpha helix \\ 
\hline
Antimicrobial LCI & 2b9k & $47$ & $0.383$ & 1 beta sheet \\ 
\hline
Tedamistat & 3ait & $74$ & $0.351$ & 2 sheets \\ 
\hline
Myoglobin & 1mbs & $153$ & $0.379$ & 8 helices \\ 
\hline
\parbox{2cm}{Asparagine\\ synthetase} & 11as & $330$ & $0.397$ & \parbox{1.5cm}{
11 helices\\ 8 sheets}\\ 
\hline
\end{tabular}%
\label{Table:proteins}%
\end{table}%

We define an energy $E(R_{g},N)$ as the ensemble average of the model
potential energy \eqref{eq:energy} over all conformations that share a
given value of $R_{g}$ (within a certain bin size). The gradient $-\partial
E/\partial R_{g}$ gives the pressure-force experienced by the protein as a
function of radius, which in principle can be observed in force spectroscopy
experiments \citep{Cecconi, Fernandez}. In this sense, we call $E(R_{g},N)$
an \textquotedblleft elastic energy\textquotedblright .

\begin{figure}[htbp]
\centering
\includegraphics[width=10cm]{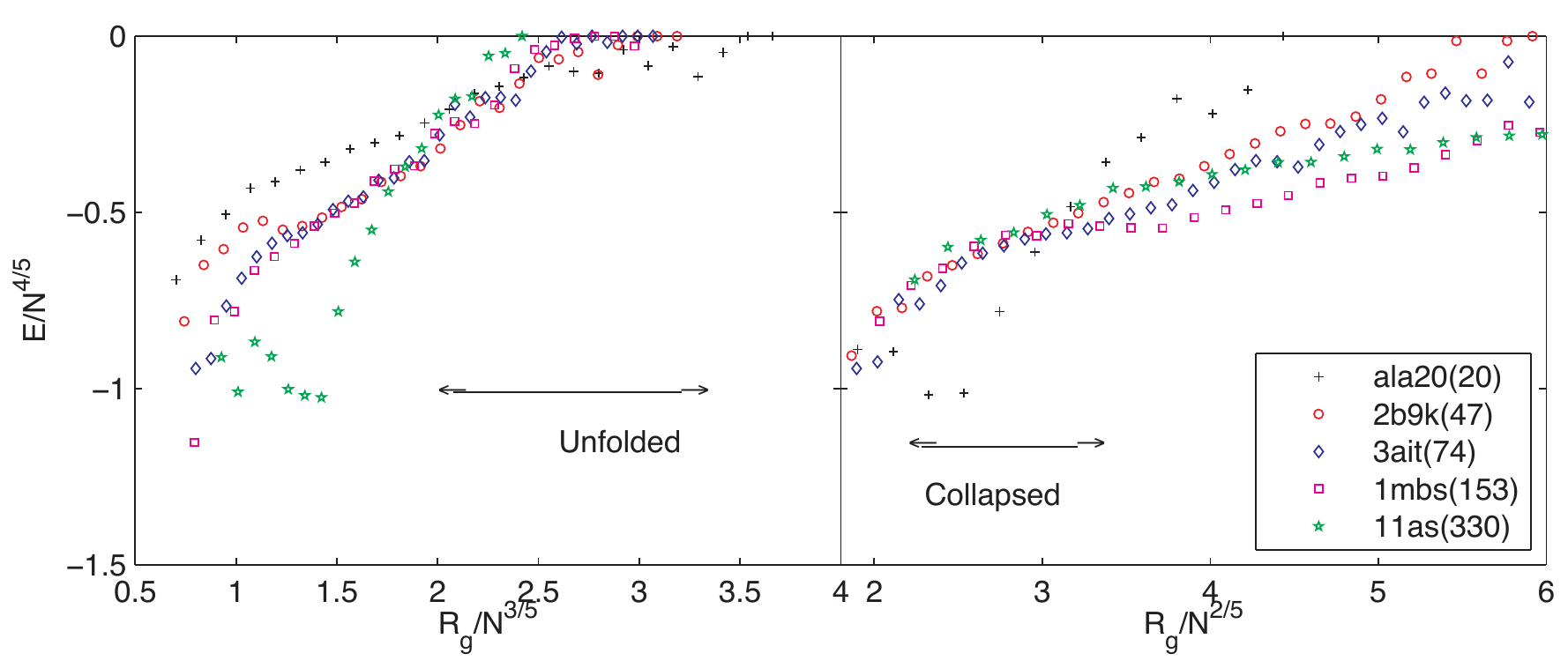}
\caption{Rescaling the data to exhibit universality in the unfolded state
(left) and collapsed state (right) of five proteins, listed with residue numbers in parenthesis. The case $\mathrm{Ala}_{20}$  is exceptional, being completely hydrophobic. Replotted from 
\protect\cite{LeiHuang:2009}}
\label{fig:es}
\end{figure}

To plot the data, we scale the energy $E$ by a factor $N^{4/5}$, as
suggested by \eqref{4/5}. Our earlier discussion indicate that the radius $%
R_{g}$ should scale like $N^{\nu }$, with $\nu =3/5$ and $\nu =2/5$ in
different regions. Accordingly we rescale the data in two different manners, as
shown in Figure \ref{fig:es}. As we can see, the 3/5 plot exhibits
universal behavior in the unfolded region, while the 2/5 plot shows
universality in the collapsed region. The protein Ala20 is exceptional,
since it is completely hydrophobic. This exception in fact shows the
relevance of hydrophobicity.

From physical arguments, we suggest the following analytical form of the
elastic energy:\footnote{%
For detailed arguments see \citet{LeiHuang:2009}.}  
\begin{equation}
E(R_{g},N)=aN^{4/5}+b(N\,R_{g})^{1/2}+c(\rho )N^{2}/R_{g}^{3}.  \label{eq:e1}
\end{equation}%
Here, $a$ and $b$ are parameters, and $c(\rho )$ represents an   
\index{effective excluded volume} effective excluded volume, which depends on $N$ and $R_{g}$
only through the scaled radius%
\begin{equation}
\rho =%
\frac{R_{g}}{N^{2/5}}.
\end{equation}
We expect $a$ and $b$ to be universal coefficients, but not the effective 
excluded volume, as it depends on the intricate structure of a collapsed
chain. 

The various terms have the following physical meaning:
\begin{itemize}
\item The first term $aN^{4/5}$merely establishes the zero point of energy
as that of a completely extended chain---a convention used in the CSAW
simulations.
\item The second term $b(N\,R_{g})^{1/2}$ represents the hydrophobic energy,
with $b$ depending on the hydrophobicity of the protein chain.
\item The term $c(\rho )N^{2}/R_{g}^{3}$ is a combination of excluded-volume energy and that from hydrogen bonding, and the relative importance of
these contributions depends on the scaled radius $\rho $.
\end{itemize}

Using a phenomenological approach, we expand $c(\rho )$ in an inverse power
series of its argument: 
\begin{equation}
c(\rho )=\sum_{n\geq 0}c_{n}\rho ^{-n}.
\end{equation}%
Terms of higher $n$ become increasingly important as the collapsed chained
is being compressed, and $c(\rho )$ can be positive or negative, depending
on the signs of various short-distance interactions.

\begin{figure}[hbtp]
\centering
\includegraphics[width=6cm]{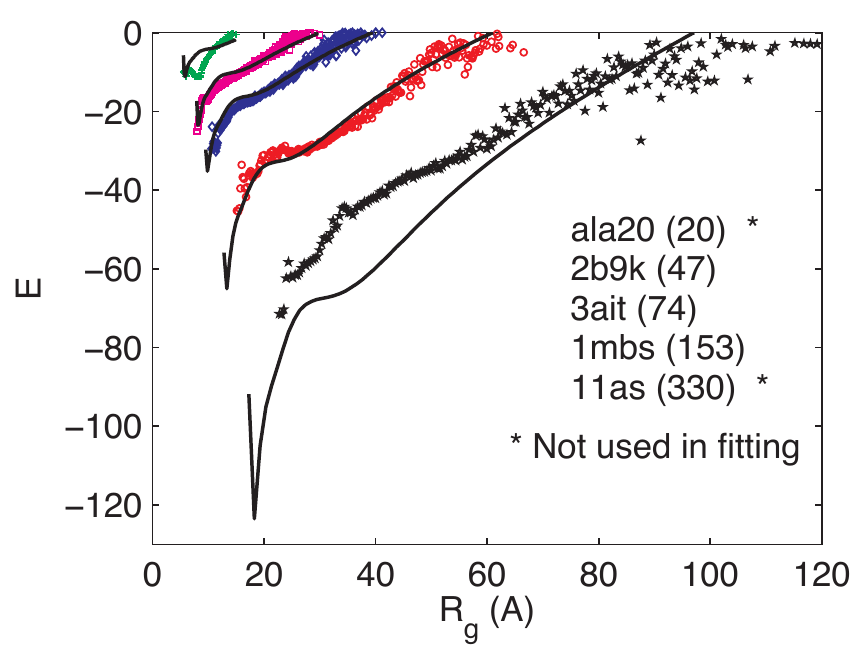}
\caption{Average potential energy \textit{vs.} radius of gyration, from computer simulations (points) and the universal elastic energy  \eqref{eq:e1} (solid curves) of five proteins, listed with residue numbers in parenthesis, in the order of the curves from top to bottom. Replotted from \citet{LeiHuang:2009}}
\label{fig:fitene}
\end{figure}

We fit the parameters $a,b,c_{n}(n=0, \cdots, 12)$ to simulation data\footnote{The non-zero parameters are: $a = -1.50$, $b=0.43$, and for even $n$ from $0$ to $12$: $c_n = 2^{n+3}\times\{2.53,-34.49,155.87,-326.50,350.09,-187.19,39.59\}$.}. Figure \ref{fig:fitene} shows the fit of \eqref{eq:e1} to the five proteins used. 
The fit is poor for $\mathrm{Ala}_{20}$ because of the exceptional nature of the protein
as remarked previously. In the case of 11as, simulation has not yet
reached an equilibrium ensemble. We did not use these two cases in fitting
the parameters. 

Thus, we have a \textquotedblleft universal\textquotedblright   elastic energy that should be
\begin{itemize}
\item truly universal for large $R_{g}$, when the short-distance interactions are not important,
and
\item valid only on average at smaller $R_{g},$ when short-distance interactions become important.
\end{itemize}

\section{Stages of protein folding}

We identify a stage by the scaling index $\nu$. Such a state is not necessarily 
stable or metastable, i.e., it does not always correspond to a local
minimum of the elastic energy \eqref{eq:e1}. But even an unstable state has
kinetic meaning, and we can identify it through physical arguments, by
imagining that certain effects are \textquotedblleft turned off\textquotedblright.

\subsection{Unfolded stage}

In the unfolded stage the natural variable is $R_{g}/N^{3/5}$, as revealed
by Figure \ref{fig:es} (left panel). In a thought experiment, we can render it stable by imagiing
that the hydrophobic effect is turned off.
imaging
\subsection{The pre-globule}

\begin{figure}[tbph]
\centering
\includegraphics[width=11cm]{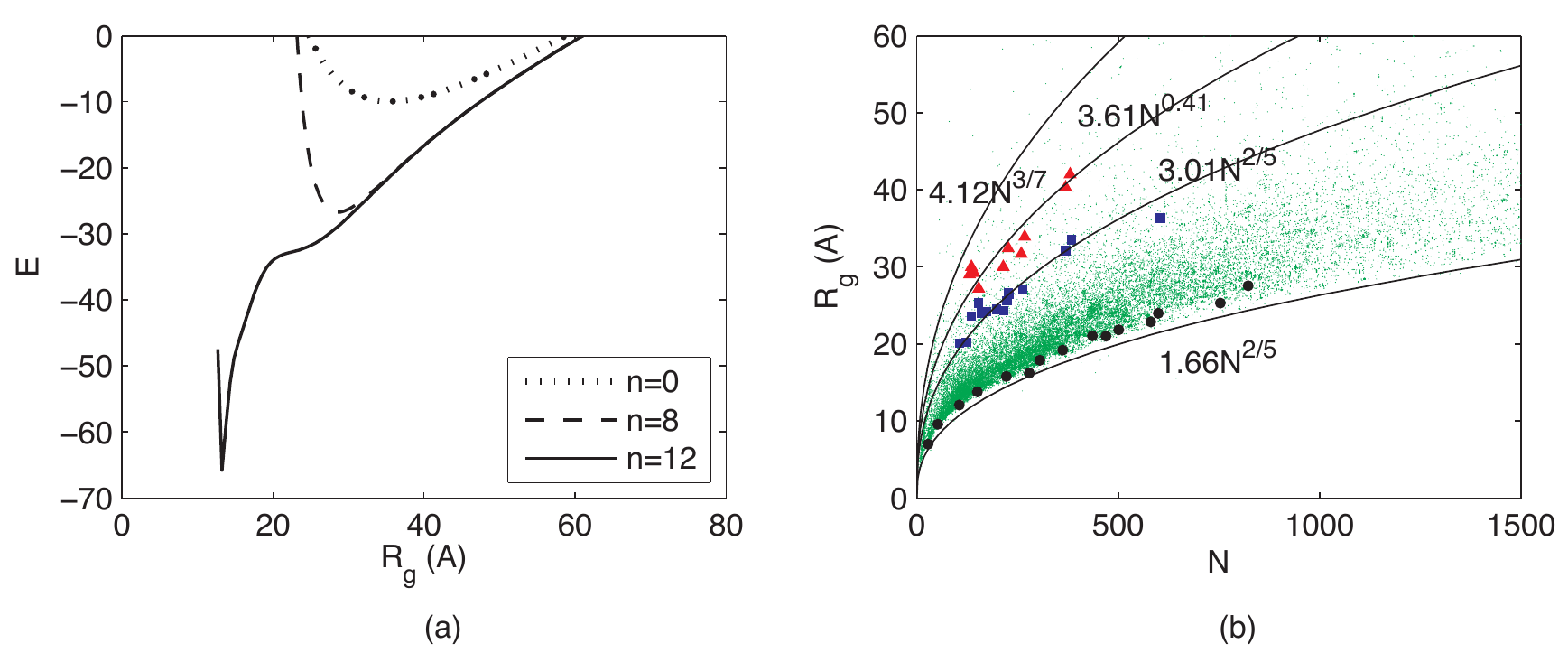}
\caption{(a). Elastic energy of 1mbs from the potential function \eqref{eq:e1} with the powers in $c(\rho)$ truncated at $n=0$ (dotted line), $n=8$ (dashed line), and $n=12$ (solid curve), respectively. (b). Radius of gyration for different stages. Theoretical pre-globule: $R_{g}=4.12N^{3/7}$; observed pre-globule: $R_{g}=3.61N^{0.41}$; molten globule: $R_{g}=3.01N^{2/5}$; the most
compact state: $R_{g}=1.66N^{2/5}$. Experimental data are also shown, the proteins in pre-globule (triangles)  and molten globule (squares) from \protect\cite{Tch2001}, the most compact proteins from \protect\cite{Arteca95} (circles), and the native proteins from
PDB (dots) \citep{Hong2009}.}
\label{fig:energy}
\end{figure}

The pre-globule is characterized by $\nu =3/7$. It may or may not be
experimentally visible, depending on details of hydrogen bonding. By
\textquotedblleft turning off\textquotedblright higher-order terms in $c( \rho) $, keeping only the $n=0$ term, we have the elastic energy
\begin{equation}
E_{1}(R_{g},N)=aN^{4/5}+b(N R_{g})^{1/2}+c_{0}N^{2}/R_{g}^{3}.
\end{equation}
This has a local minimum with scaling law $R_{g}\sim N^{3/7}$ (Figure \ref{fig:energy}), which defines the ideal pre-globule theoretically.

Figure \ref{fig:energy}(b) shows the radius of the ideal pre-globule, which
is expected to scale according to 
\begin{equation}
R_{g}\approx 4.12N^{3/7} \quad (\mbox{theoretical\ pre-globule}).
\end{equation}
This theoretical prediction is larger than observed data, because the
hydrogen-bonding attractions have been ignored. By extending the  powers in $%
c(\rho )$ to $n=8$, we find that the local minimum satisfies the scaling law 
\begin{equation}
R_{g}\approx 3.61N^{0.41}\quad (\mbox{observed\ pre-globule}),
\end{equation}
 in good agreement with the pre-globule observed in
some proteins (Figure \ref{fig:energy}) \citep{Uversky02}.

The higher powers in $c\left( \rho \right) $ destabilize the pre-globule, due
to the fact that the attractive contributions from hydrogen bonding may make 
$c(\rho )$ change sign. 

\textquotedblleft On average\textquotedblright, therefore, the pre-globule is not even metastable, and the
collapsed protein chain continues to shrink until it reaches the most
compact state, with scaling law $R_{g}\sim N^{2/5}$.

\subsection{The molten globule}

In the energy function \eqref{eq:e1} as plotted in Figure \ref{fig:energy}(a) (solid line), there is a flat
shoulder, which is almost metastable. Between this shoulder and the lowest
minimum, corresponding to the most compact state, there is a region that can
be identified as the molten globule. We identify this region with a single
stage because the scaling exponent is 2/5 throughout. It is bounded by the curves plotted in Figure \ref{fig:energy}(b):
\begin{equation}
\label{eq:rmm}
\begin{array}{ll}
R_{g} \approx 3.01N^{2/5}& (\mathrm{beginning\ of\ molten\ globule}),
\\
R_{g} \approx 1.66N^{2/5}& (\mathrm{most\ compact\ state}).
\end{array}%
\end{equation}
As shown in Figure \ref{fig:energy}(b), data from the PDB show that the $R_{g}$
of all native proteins are distributed between these two curves.

\subsection{The native state}

The most compact state is the closest we can get to the native state in
CSAW, since the detailed structures of the side chains have not been
included. As Figure \ref{fig:energy}(b) shows, the native state in real proteins shares the same scaling index
2/5 with the molten globule. While the time it takes for a random coil to
collapse into the molten globule is less than or of the order of
microseconds, the development of the molten globule to the native state
requires seconds or longer. This suggests that after the molten globule, the
\textquotedblleft snapping on\textquotedblright of side chains is governed by new mechanisms.

\subsection{Mechanisms of different stages}

In summary, as shown in Figure \ref{fig:stages},
\begin{itemize}
\item The unfolded state is characterized by an absence of both hydrophobic forces
and hydrogen bonding. The dominant interaction is the excluded-volume effect,
which makes the unfolded protein chain a SAW.
\item Under hydrophobic forces, the protein chain rapidly collapses into the
pre-globule, which is maintained by a balance between the hydrophobic
pressure and excluded-volume effect. The density is not high enough for
significant hydrogen bonding to occur, and the chain can slide against
itself. The lifetime of this state depends on the importance of
hydrogen bonding.
\item Upon further compression, hydrogen bonding occurs, and the structure
acquires rigidity. The protein is now essentially as compact as the native
state.
\item The side chains \textquotedblleft snap on\textquotedblright to native positions, and this process takes a
macroscopically long time.
\end{itemize}

\section{Phase transition in myoglobin}

\begin{figure}[htbp]
\centering
\includegraphics[width=6cm]{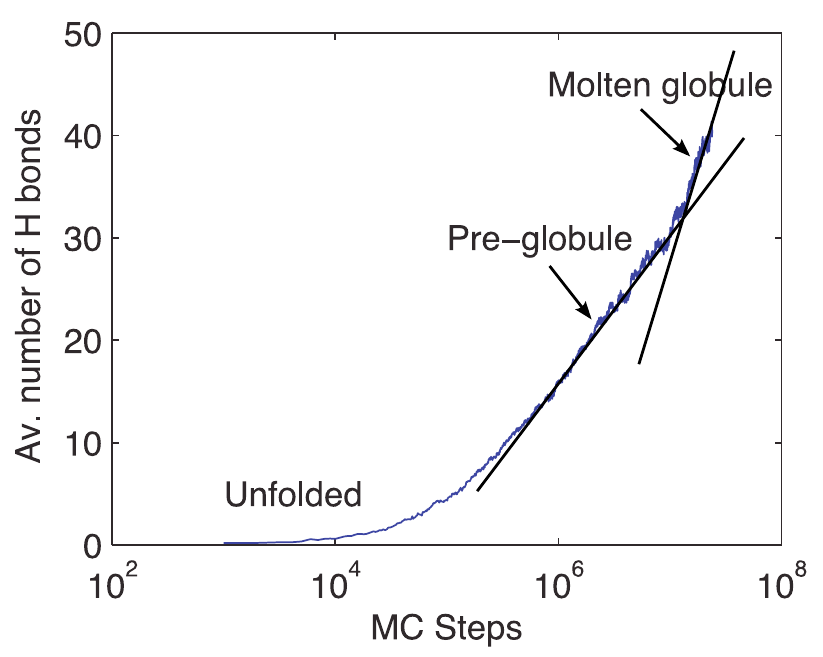}
\caption{Growth of hydrogen bonds in the simulation of myoglobin. The onset of the transition from pre-globule to molten globule is marked by a sudden jump in the growth rate. Replotted from \citet{LeiHuang:2009}}
\label{fig:trasmyo}
\end{figure}

The transition from pre-globule to molten globule shows up as a sudden
increase in the rate of hydrogen bonding, in a process akin to a phase
transition. We illustrate this in the case of myoglobin, as shown in 
Figure \ref{fig:trasmyo} with the number of hydrogen bonds as a function of time.

\section{A theoretical limit on protein size}

To describe the effective excluded-volume effect through a single variable depending
only on the residue number $N$, we define 
\index{excluded volume} the \textquotedblleft excluded volume\textquotedblright $%
\bar{c}$ as the average of $c(\rho )$ over all physically possible values of 
$\rho $: 
\begin{equation}
\bar{c}=\dfrac{1}{\rho _{\max }-\rho _{\min }}\int_{\rho _{\min }}^{\rho
_{\max }}c(\rho )d\rho ,  \label{eq:ev}
\end{equation}%
where $\rho _{\min }$ and $\rho _{\max }$ are respectively the minimum and
maximum values of the scaled radius $\rho = R_g/N^{2/5} $ in the ensemble.

\begin{figure}[htbp]
\centering
\includegraphics[width=6cm]{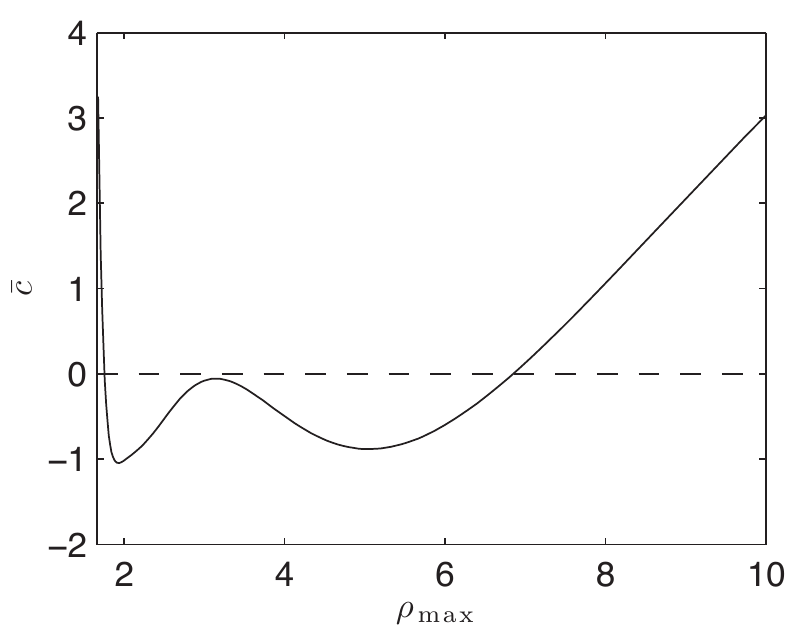}
\caption{The excluded volume $\bar{c}$ as a function of $\rho_{\max}$.}
\label{fig:cbar}
\end{figure}

We have $\rho _{\min }=1.66$ according to the radius of the most compact state given by \eqref{eq:rmm}.  From Figure \ref{fig:energy}(b),  when there is no hydrogen bonding, the chain
collapses to a state with scaling law $R_{g}\approx 4.12N^{3/7}$. This can
serve as the maximal radius in the post-collapsed ensemble, and thus
\begin{equation}
\label{eq:rmax}
\rho_{\max }\approx 4.12N^{1/35}.
\end{equation}
Figure \ref{fig:cbar} shows $\bar{c}$ is a
function of $\rho _{\max }$.

From Figure \ref{fig:cbar}, we see that $\bar{c}$ changes from negative to positive at $%
\rho_{\max}\approx 6.85$. When $4.12<\rho_{\max}<6.85$, the 
excluded volume is negative, and therefore attractive interactions are
dominant, which drive the chain to a compact state. When $%
\rho_{\max}>6.85$, however, the excluded volume is positive, and
repulsive interactions are dominant. In this case, the chain cannot fold to a more compact state, because not enough
hydrogen bonds can be formed after the hydrophobic collapse.

The value $\rho _{\max }=6.85$ and \eqref{eq:rmax} yield a critical chain length 
\index{critical length} 
\begin{equation}
N_{\mathrm{crit}}=5.34\times 10^{7}.
\end{equation}
Thus, a super protein ($N>N_{\mathrm{%
crit}}$) do not have a stable folded state.

We emphasize here that the above conclusion is valid only for a natural protein, i.e., the fraction of hydrophobic residues is around 40\% %
\citep{Hong2009}. It may not be the case for a fully hydrophobic chain.

\section{Discussion and outlook}

In this chapter, we discuss protein folding from the perspective of
statistical physics. We treat the protein as a physical chain immersed in water,
and tends towards thermodynamic equilibrium. The process is stochastic, and
can be modeled with the master or the Langevin equation. We introduce the CSAW model (conditioned self-avoiding
walk), which simulates the Langevin equation for the protein chain. With this model, we are able to study various aspects of protein folding. 

In treating protein folding as a physical process, the CSAW model differs
from MD in two important aspects, namely
\begin{itemize}
\item irrelevant degrees of freedom are ignored; and
\item the environment is treated as a stochastic medium.
\end{itemize}
These, together with simplifying treatment of interactions, enable the model
to produce qualitatively correct results with minimal demands on computer
time.

An important simplification is the separation of the hydrophobic effect and
hydrogen bonding, as expressed by the separate terms in the potential
energy. Since both effects arise physically from hydrogen bonding, it is not
obvious that we can make such a separation. The implicit assumption is that
hydrogen bonding with water involves only the side chains, while internal
hydrogen bonding involves only the backbone. This property is supported by
statistical data, but should be a result rather an assumption of the model.
We should try to remedy this in an improved version of the model.

The CSAW model successfully folds polyalanine ($\mathrm{Ala}_{20}$), a
protein fragment of 20 identical amino acids alanine, from random coil to
its native state, which is a single alpha helix \citep{LeiHuang:2008}. We also use
the CSAW model to study the elastic energy of proteins.
The elastic energy gives rise to scaling relations of the form $R_g\sim
N^{\nu}$ in different stages of folding, with $\nu = 3/5, 3/7, 2/5$ for the unfolded
stage, the pre-globule, and the molten globule, respectively. The theoretical
predictions agree well with experimental data (Figure \ref{fig:energy}). These stages, together with mechanisms responsible for their formation, are summarized in Figure \ref{fig:stages}.

These examples show that, despite its simplicity,
CSAW incorporates important physical principles governing protein folding. 

We can refine CSAW by adding refinements, including the following:
\begin{itemize}
\item All-atom side chains with fractional
hydrophobicity \citep{Sun07}.
\item Electrostatic interactions.
\item Replacement of hard-sphere repulsions by Lennard-Jones
potentials.
\end{itemize}

We hope to make progress on the protein folding problem with this
theoretical laboratory.

\label{lastpage-01}

\bigskip

%\printindex

\end{document}